\documentclass[12pt]{article}
\usepackage{amssymb,amsmath}
\usepackage[noblocks]{authblk}
\usepackage[top=0.75in, bottom=0.75in, left=0.75in, right=0.75in, dvips]{geometry}
\usepackage{caption}
\usepackage{graphicx}
\usepackage{epstopdf}
\begin{document}
\textwidth 10.0in
\textheight 9.0in
\topmargin -0.60in
\title{Renormalization Mass Scale and Scheme Dependence in the Perturbative Contribution to Inclusive Semileptonic $b$  Decays}
\author[1]{F.A. Chishtie}
\author[2,3]{D.G.C. McKeon}
\author[4]{T.N. Sherry}
\affil[1]{Department of Applied Mathematics, The University of Western Ontario, \newline London, ON N6A 5B7, Canada}
\affil[2] {Department of Mathematics and
Computer Science, Algoma University,\newline Sault Ste. Marie, ON P6A
2G4, Canada}
\affil[3] {School of Mathematics, Statistics and Applied Mathematics, National University of Ireland Galway, University Road, Galway, Ireland H91 TK33} 
\maketitle                              

\maketitle
\noindent
PACS No.: 11.10Hi\\
Key Words: renormalization scheme, b-quark decays\\
email: fachisht@uwo.ca, dgmckeo2@uwo.ca, tom.sherry@nuigalway.ie

\begin{abstract}
We examine the perturbative calculation of the inclusive semi-leptonic decay rate $\Gamma$ for the $b$-quark, using mass-independent renormalization. To finite order of perturbation theory the series for $\Gamma$ will depend on the unphysical renormalization scale parameter $\mu$ and on the particular choice of mass-independent renormalization scheme; these dependencies will only be removed after summing the series to all orders. In this paper we show that all explicit $\mu$-dependence of $\Gamma$, through powers of ln$(\mu)$, can be summed by using the renormalization group equation. We then find that this explicit $\mu$-dependence can be combined together with the implicit $\mu$-dependence of $\Gamma$ (through powers of both the running coupling $a(\mu)$ and the running $b$-quark mass $m(\mu)$) to yield a $\mu$-independent perturbative expansion for $\Gamma$ in terms of $a(\mu)$ and $m(\mu)$ both evaluated at a renormalization scheme independent mass scale $I\!\!M$ which is fixed in terms of either the ``$\overline{MS}$ mass'' $\overline{m}_b$ of the $b$ quark or its pole mass $m_{pole}$. At finite order the resulting perturbative expansion retains a degree of arbitrariness associated with the particular choice of mass-independent renormalization scheme. We use the coefficients $c_i$ and $g_i$ of the perturbative expansions of the renormalization group functions $\beta(a)$ and $\gamma(a)$, associated with $a(\mu)$ and $m(\mu)$ respectively, to characterize the remaining renormalization scheme arbitrariness of $\Gamma$. We further show that all terms in the expansion of $\Gamma$ can be written in terms of the $c_i$ and $g_i$ coefficients and a set of renormalization scheme independent parameters $\tau_i$. A second set of renormalization scheme independent parameters $\sigma_i$ is shown to play a very similar role in the perturbative expansion of $m_{pole}$ in terms of $m(\mu)$ and $a(\mu)$. We illustrate our approach by a perturbative computation of $\Gamma$ using the $\overline{MS}$ renormalization scheme.  Two other particular mass independent renormalization schemes are briefly considered. 
\end{abstract}

\section{Introduction}

Mass independent renormalization schemes [1,2] are relatively easy to implement, but are often viewed as being ``unphysical''.  This is particularly true when the process being considered involves a heavy particle such as a $b$ quark.  In addition to this problem, perturbative results obtained using a mass independent renormalization scheme (RS) depend on a non-physical scale parameter $\mu$. Furthermore, results at a finite order of perturbation theory can be altered by making a finite renormalization of the quantities that characterize the theory (masses, couplings and field strengths). 

An analysis of the ambiguities associated with $\mu$ and the ambiguities in RS appear in ref. [8]. In this reference, it was proposed that the ambiguities could be resolved at finite order by applying the ``Principle of Minimal Sensitivity". In ref. [3] the renormalization group (RG) equation is used to sum logarithmic corrections to various processes. One can use the RG functions at $n$-loop order to sum the $N^{n-1}LL$ corrections (leading-log, next-to-leading-log, etc.).  Alternatively, one can sum all logarithmic corrections in terms of the log-independent corrections and the RG functions.  This latter approach was used in refs. [4,5] to examine $R_{e^+e^-}$, the cross section for $e^+e^- \rightarrow$ hadrons and further processes were studied in [6].  It was found that upon summing all logarithmic contributions to $R_{e^+e^-}$ the explicit dependence of $R_{e^+e^-}$ on $\mu$ cancelled with its implicit dependence on $\mu$, leaving $R_{e^+e^-}$ dependent only on the ratio $Q/\Lambda$ where $Q$ is the centre of mass energy and $\Lambda$ is a mass scale associated with the boundary value of the running coupling $a(\ln \mu /\Lambda)$.  A set of RS invariant parameters $\tau_i$ was found, and $R_{e^+e^-}$ was seen to be expressible in terms of $a(\ln Q/\Lambda)$ and $\tau_i$. The parameters $\tau_i$ are related to the RS invariant parameters found in refs. [8,25]. 

In this paper we examine the decay rate $\Gamma$ relevant for both the processes $b \rightarrow u\ell^-\overline{\nu}_\ell$ and $b \rightarrow c\ell^-\overline{\nu}_\ell$. These decays have both perturbative and non-perturbative contributions, and in this work we will restrict our attention to the ambiguities inherent in perturbative calculations involving $b$ quark decay. Addressing such an issue involves a massive parameter (the mass of the $b$ quark) contributing explicitly to $\Gamma$ which leads to a complication in the analysis as when using a ``mass independent'' RS, the mass now ``runs''; this is different from the computation of $R_{e^+e^-}$ where there is a fixed massive scale $Q$, the centre of mass energy, entering the calculation.  (A calculation of $\Gamma$ which uses the pole mass [16] of the $b$ quark [3,7] rather than its running mass $m(\mu)$ is possible, but such a calculation does not use a mass independent RS.)

It is possible to use the RG approach to carry out various partial summations of the perturbative expansion of the decay rate $\Gamma$. The first of these involves use of the RG equation to sum the $LL, NLL, \ldots N^pLL$ contributions to $\Gamma$ in terms of the $1, 2,\ldots, {(p+1)}$ loop contributions to $\beta$ and $\gamma$, the RG functions associated with $a(\mu)$ and $m(\mu)$ respectively.  A second approach is to sum all logarithmic contributions to $\Gamma$ so that $\Gamma$ is expressed in terms of its log independent part. One appealing feature of this  log-summed form is that the explicit and implicit dependence on $\mu$ is again shown to cancel, much as it did in the analysis of $R_{e^+e^-}$.  This removes one source of ambiguity in any subsequent calculation of the decay rate $\Gamma$.

Even within the mass independent renormalization schemes, there is a degree of RS ambiguity beyond the ambiguities arising from $\mu$.  The renormalization scheme can be parameterized by the appropriate coefficients $c_i$ and $g_i$ in the loop expansion of the RG functions associated with the coupling constant and the anomalous mass dimension respectively [8,9].  The requirement that $\Gamma$ be RS independent leads to a set of RS invariant parameters $\tau_i$.  Different RS's are considered, each leaving $\Gamma$ expressed in terms of these $\tau_i$, a mass parameter $I\!\!M$ associated with the $b$ quark, and a QCD scale parameter $\Lambda$ as well as the parameters $c_i$ and $g_i$ appropriate for these $RS$'s.  In one of these schemes, the perturbative expansion for $\Gamma$ terminates after a finite number of terms, while in the other ('t Hooft-like scheme [10]) the perturbative expansions for the RG functions $\beta$ and $\gamma$ terminate.

We also apply RG summation to relate the pole mass [16] for the $b$ quark $m_{\rm{pole}}$ to its running mass $m(\mu)$.  We use this relation to show how $\Gamma$ can be found in terms of the physical, RS invariant mass scale, $m_{pole}$ in place of the mass parameter $I\!\!M$.

\section{Renormalization Group Summation}

A perturbative evaluation of the amplitude $\Gamma$ for the semi-leptonic decay process $b \rightarrow u\ell^-\overline{\nu}_\ell$ or $b \rightarrow c \ell^-\overline{\nu}_\ell$ leads to the expression\footnote{The expression for $\Gamma$ in eq. (1) is distinct from the way the solution to the RG equation of eq. (3) is usually presented [11] as it does not involve simply replacing the mass and coupling in a perturbative expression for $\Gamma$ by the running coupling and mass. This is discussed in more detail in ref. [12]. Eq. (1) is an ansatz for the solution to the RG equation that is motivated by knowing the form of results obtained from evaluating the $n$-loop Feynman diagrams needed to obtain the coefficients $T_{n}(k=0,1..n)$.}
\begin{equation}
\Gamma_q  = \frac{G_F^2|V_{qb}|^2}{192\pi^3}[ m(\mu)]^5 \sum_{n=0}^\infty \sum_{k=0}^n T_{n,k} a^n(\mu) \ln^k \left(\frac{\mu}{m(\mu)}\right)
\end{equation}
where $m(\mu)$ is the running mass for the $b$ quark, $a(\mu)$($=\alpha_s(\mu)/\pi$) is the strong coupling, $q = c$ or $u$ as appropriate, and we assume five active flavours.  We will work with the reduced decay width
\begin{equation}
\Gamma = \Gamma_q \left( \frac{G_F^2|V_{qb}|^2}{192\pi^3}\right)^{-1}
\end{equation}
throughout this paper. As the exact expression for $\Gamma$ is independent of the unphysical renormalization scale parameter $\mu$ we have the RG equation
\begin{equation}
\mu \frac{d\Gamma}{d\mu} = 0 = \left(\mu \frac{\partial}{\partial\mu} + \beta(a) \frac{\partial}{\partial a} + m\gamma(a) \frac{\partial}{\partial m}\right)\Gamma
\end{equation}
where
\begin{equation}
\beta(a) = \mu \frac{\partial a}{\partial\mu} = -ba^2 (1 + ca + c_2 a^2 + \ldots)
\end{equation}
and
\begin{equation}
m\gamma(a) = \mu \frac{\partial m}{\partial\mu} = mfa (1 + g_1 a + g_2 a^2 + \ldots).
\end{equation}

Eq. (1) ensures that the explicit dependence on $\Gamma$ on $\mu$ through $\ln(\mu)$ and its implicit dependence through $a(\mu)$ and $m(\mu)$ cancels. Approximating the computation of $\Gamma$ up to, and including, $N$-loop Feynman diagrams involves truncating the summation process in eq. (1)
\begin{equation}
\Gamma^{(N)} (\mu) = [m(\mu)]^5 \sum_{n=0}^N \sum_{k=0}^n T_{n,k} a^n(\mu)\ln^k\left( \frac{\mu}{m(\mu)}\right).
\end{equation}
This $N$-loop approximation is highly dependent on the arbitrary renormalization scale $\mu$, both explicitly and implicitly, as illustrated in ref. [3] for $N = 3$.  No convincing argument for specifying a particular value of $\mu$ in $\Gamma^{(N)}(\mu)$ has gained acceptance.

In ref. [3] we also considered a re-ordering of the summation in eq. (1) so that eq. (3) could be used to sum the contributions to $\Gamma$ of all leading-logarithms $(LL)$, next-to-leading-logarithms $(NLL)$ etc., which led to a lessening of the $\mu$-dependence in a finite order approximation; however, this approach did not totally eliminate dependence on $\mu$.  The $N^PLL$ sum contributing to $\Gamma$ receives contributions from all $T_{p+k,k}(k \geq 0)$ with $T_{p,0}$ providing a boundary value, and from the $\beta$ and $\gamma$ functions up to order $p + 1$ in the loop expansion.

In this paper we use a different way of organizing the sums in eq. (1).  In this approach, we will sum all log dependent contributions to $\Gamma$.  The RG equation will make it possible to express all these log dependent contributions in terms of log independent parts in such a way that all dependence on $\mu$ is completely removed.  This approach is used in the discussion of $R_{e^+e^-}$ in refs. [4,5].  However, the summation is more complicated here than it was for $R_{e^+e^-}$.  This is because eq. (1) shows that $\Gamma$ does not directly depend on a physical mass parameter which would be analogous to the centre of mass energy $Q$ that occurs in refs. [4,5].  Instead a ``running mass'' $m(\mu)$ associated with the renormalized mass of the $b$ quark appears.  For $R_{e^+e^-}$ one can very simply sum all of the log-dependent contributions to show immediately that all $\mu$ dependence is removed.  This result is also possible for $\Gamma$, but it proves to be more awkward to sum all of its log-dependent pieces.

We begin by defining
\begin{equation}
A_n(a(\mu)) = \sum_{k=0}^\infty T_{n+k,n} a(\mu)^{n+k}
\end{equation}
so that eq. (1) leads to
\begin{equation}
\Gamma = m^5(\mu) \sum_{n=0}^\infty A_n(a(\mu))\ell^n
\end{equation}
where now $\ell = \ln(\mu/m(\mu))$.  We recognize $A_n(a(\mu))$ in eq. (7) as the sum of all the coefficients of $\ell^n$ in the perturbative expansion of $\Gamma$ in eq. (1).  In particular, $A_0(a(\mu))$ contains all of the log independent parts of $\Gamma$. If eq. (8) is substituted into eq. (3) we find
\begin{equation}
A_n(a(\mu)) = \frac{-1}{n}\left[ \hat{\beta}(a(\mu)) \frac{\partial}{\partial a} + 5 \hat{\gamma} (a(\mu))\right] A_{n-1} (a(\mu))
\end{equation}
where
\begin{subequations}
\begin{align}
 \hat{\beta} &= \beta/(1-\gamma), \\
 \hat{\gamma} &= \gamma/(1-\gamma).
\end{align}
\end{subequations}

We now define
\begin{equation}
E(a(\mu)) = \exp \left[\int_0^{a(\mu)} dx \frac{\gamma(x)}{\beta(x)} + \int_0^K dx \frac{fx}{bx^2(1+cx)}\right]
\end{equation}
where $K$ is some cut off. The second integral in eq. (11) is a RS independent infinite constant whose role is to ensure that the argument of the exponential is finite.

We also note that solutions to eqs. (4,5) can be written as
\begin{equation}
\ln \left( \frac{\mu}{\Lambda}\right) = \int_0^{a(\mu)} \frac{dx}{\beta(x)} + \int^K_0 \frac{dx}{bx^2(1 + cx)}
\end{equation}
and
\begin{equation}
m(\mu) = I\!\!M E(a(\mu))
\end{equation}
where $\Lambda$ and $I\!\!M$ are scale dependent quantities used to define boundary conditions on eqs. (4,5).  We note that in eqs. (12,13) a change in $K$ can be absorbed into changes in $\Lambda$ and $I\!\!M$.  The scale parameter $I\!\!M$ is closely related to a scalar parameter that is RS invariant that was previously used in ref. [26]. In refs. [8,9], $K$ is taken to be infinite.  (Below we will often use $a(\mu)$ to denote $a\left( \ln \frac{\mu}{\Lambda}\right)$. Similarly, the $\mu$ dependence of any dimensionless quantity, such as $E$, will be written $E(a(\mu))$ but will be understood to mean $E(a(\ln\mu/\Lambda))$. ) 

We now can re-express eq. (9) in the form
\begin{equation}
B_n(a(\mu)) = \frac{-1}{n} \hat{\beta} (a(\mu)) \frac{\partial}{\partial a} B_{n-1} (a(\mu))
\end{equation}
where
\begin{equation}
B_n(a(\mu)) = E^5(a(\mu))A_n(a(\mu)).
\end{equation}
We now define an auxiliary quantity $\eta$ so that
\begin{equation}
\frac{\partial}{\partial \eta} = \hat{\beta}(a) \frac{\partial}{\partial a}
\end{equation}
so that by eq. (10a)
\begin{equation}
\eta (a(\mu)) = \int_0^{a(\mu)} dx \frac{1-\gamma(x)}{\beta(x)} + \int_0^K dx \frac{1-fx}{bx^2(1+cx)}.
\end{equation}
Together eqs. (14,16) show that
\begin{equation}
B_n (a(\mu))  = \frac{-1}{n}\frac{\partial}{\partial \eta} B_{n-1} (a(\mu))
\end{equation}
or upon iteration,
\begin{equation}
= \frac{(-1)^n}{n!} \frac{\partial^n}{\partial \eta^n} B_0 (a(\mu)).
\end{equation}
 
Eqs. (8,15,19) together lead to
\begin{equation}
\Gamma = m^5(\mu) E^{-5}(a(\mu)) \sum_{n=0}^\infty \frac{(-\ell)^n}{n!} \left(\frac{\partial}{\partial \eta}\right)^n B_0 (a(\mu)).
\end{equation}
Eq. (13) results in eq. (20) becoming
\begin{equation}
\Gamma = I\!\!M^5 B_0(a(\eta - \ell)).
\end{equation}
However, we now see by eqs. (11-13,17), that
\begin{equation}
\eta - \ell = \ln \left( \frac{I\!\!M}{\Lambda}\right)
\end{equation}
which is $\mu$ independent.  We now can combine eqs. (15,21,22) to obtain
\begin{subequations}
\begin{align}
\Gamma = I\!\!M^5 E^5 \left( \ln \frac{I\!\!M}{\Lambda}\right) A_0 \left( a\left( \ln \frac{I\!\!M}{\Lambda}\right)\right) \\
= m^5 \left( \ln \frac{I\!\!M}{\Lambda}\right) A_0 \left(a\left( \ln \frac{I\!\!M}{\Lambda}\right)\right)
\end{align}
\end{subequations}
We see that $\Gamma$ is now expressed in terms of its log-independent contribution $A_0$ and that it is $\mu$ independent.  Even when we are restricted to a finite number of terms in the expansion of $A_0\left( a \left( \ln \frac{I\!\!M}{\Lambda}\right)\right)$,
\begin{equation}
\sum_{k=0}^N T_{k,0} a\left(\ln \frac{I\!\!M}{\Lambda}\right)^k
\end{equation}
the resulting approximation to $\Gamma$ is still $\mu$-independent.  However it retains a residual RS dependence as will be discussed in the following section.  It is interesting that eq. (23) can be obtained from eq. (1) by dropping all terms involving $\left(\ln \frac{\mu}{\Lambda}\right)^k(k \geq 1)$ and then setting $\mu = I\!\!M$.  The exact expression for the mass scale $I\!\!M$ appearing throughout eq. (23) will be shown to be a RS independent (though unphysical) quantity.  However, in section four it will be shown that $I\!\!M$ can be expressed in terms of the pole mass of the $b$ quark, $m_{pole}$, a physical quantity that is RS invariant.

In eq. (23b), $A_0\left( a \left( \ln \frac{I\!\!M}{\Lambda}\right)\right)$ is to be evaluated by computing Feynman diagrams and is expressed as a power series in $a \left( \ln \frac{I\!\!M}{\Lambda}\right)$.  The convergence of such series is not clearly understood; it is generally accepted that they are asymptotic series and that divergences due to ``renormalons'' [10,13] can arise.  The problem of convergence of the perturbative series in $a$ is not a focus of our work; our focus is on the ambiguities that arise to a finite order in a perturbative expansion when using mass independent renormalization.  We have shown that the ambiguity residing in eq. (1) due to the presence of $\mu$ when this series is truncated can be fixed by using the RG equation to sum those logarithms containing explicit dependence on $\mu$.  Other approaches to eliminating ambiguities arising in the computation of perturbative effects are "The Principle of Maximal Conformality" (PMC) [27,28] and "Complete Renormalization Group Improvement" (CORGI) [29,31]. A comparison of CORGI with the approach used here appears in [30]. We now consider ambiguities residing in the choice of RS.

\section{Renormalization Scheme Dependence}

It has been established that the RS ambiguities within mass independent RS can be parameterized by the mass scale $\mu$ as well as the coefficients $c_i(i \geq 2)$, $g_i(i \geq 1)$ in eqs. (4,5).  In a recent paper [14], we examined how the running coupling $a(\mu)$ and the running mass $m(\mu)$ vary with RS.

Focussing first on $\mu$, the scale dependence, we found
\begin{subequations}
\begin{align}
a(\mu) &= a(\mu^\prime) \left[1+b\lambda a(\mu^\prime) + b\lambda (c + b\lambda)a^2(\mu^\prime) + b\lambda\left( c_2 + \frac{5}{2} bc\lambda + b^2\lambda^2\right)a^3(\mu^\prime) + \ldots \right]\\
m(\mu) &= m(\mu^\prime) \left[1-f\lambda a(\mu^\prime) - f\lambda \left( g_1 - \frac{1}{2}(f-b)\lambda\right)a^2(\mu^\prime)\right. \nonumber \\
 & - f\lambda
 \left. \left(g_2+ \Big( bg_1 + \frac{bc}{2} - fg_1\Big)\lambda + \frac{1}{3}(f-b)\Big(\frac{f}{2}-b\Big)\lambda^2\right)a^3(\mu^\prime) + \ldots \right] 
\end{align}
\end{subequations}
where $\lambda = \ln(\mu^\prime/\mu)$.  These results tell us how the running coupling and running mass at two different scales are related, all within one RS.

Turning now to the dependence on the $c_i$ and $g_i$ parameters, it was shown in refs. [8,9] that
\begin{subequations}
\begin{align}
\frac{\partial a}{\partial c_i}&= B_i(a)\nonumber \\
&= -b\beta(a) \int_0^a dx \frac{x^{i+2}}{\beta^2(x)}\nonumber \\
&= a^{i+1} \left( W_0^i + W_1^i a+ W_2^i a^2 + \ldots \right)\\
\frac{1}{m} \frac{\partial m}{\partial c_i} &= \Gamma_i^c(a)\nonumber \\
&= \frac{\gamma(a)}{\beta(a)}B_i(a) + b \int_0^a dx \frac{x^{i+2}\gamma(x)}{\beta^2(x)}\nonumber \\
&= a^i \left( U_0^i + U_1^i a+ U_2^i a^2 + \ldots \right)\\
\frac{1}{m} \frac{\partial m}{\partial g_i} &= \Gamma_i^g(a)\nonumber \\
&= f \int_0^a dx \frac{x^{i+1}}{\beta(x)}\nonumber \\
&= a^i\left( V_0^i + V_1^i a + V_2^i a^2 + \ldots \right)\\
\intertext{and}
\frac{\partial a}{\partial g_i} &= 0.
\end{align}
\end{subequations}
The first few $W_j^i$, $U_j^i$ and $V_j^i$ coefficients can be found in ref. [14].  We have also derived the relationship between the running couplings $a(\mu , c_i)$ and the running masses $m(\mu , c_i, g_i)$ at the same renormalization scale but at different values of the parameters $c_i$ and $g_i$.  If
\begin{align}
a^\prime & = a(\mu , c_i^\prime) , a = a(\mu , c_i)\nonumber \\
m^\prime & = m(\mu , c_i^\prime , g_i^\prime) , m = m(\mu , c_i, g_i)\nonumber
\end{align}
then the first few terms in these relationships are ($\rho\equiv\frac{f}{b}$)
\begin{subequations}
\begin{align}
a^\prime &= a\big[ 1 + \left(c_2^\prime - c_2\right) a^2 + \frac{1}{2}\left( c_3^\prime - c_3\right) a^3 + \ldots \big]\\
m^\prime &= m \big[ 1 + \rho \left(g_1 - g_1^\prime\right) a + \frac{\rho}{2}\big(g_2 - g_2^\prime  + c_2 - c^\prime_2 \nonumber\\
    & \hspace{2cm}- c  \left(g_1 - g_1^\prime\right) + \rho  \left(g_1 - g_1^\prime\right)^2\big)  a^2 + \ldots \big]
\end{align}
\end{subequations}
Further terms in these relationships can be found in ref. [14].

In ref. [14] we also reorganized the summations in eq. (25) to end up with
\begin{subequations}
\begin{align}
a(\mu) &= \sum_{n=0}^\infty X_n (a(\mu^\prime)\lambda) a(\mu^\prime)^{n+1}\\
m(\mu) &= m(\mu^\prime) \sum_{n=0}^\infty Y_n (a(\mu^\prime)\lambda) a(\mu^\prime)^n
\end{align}
\end{subequations}
where, as before, $\lambda = \ln(\mu^\prime/\mu)$.  $X_n$ and $Y_n$ have been computed iteratively up to $n = 4$ corresponding to the sums of all leading-logarithms, next-to-leading-logarithms $\ldots$ (next-to)$^4$-leading-logarithms.  In particular, the first three solutions are
\begin{subequations}
\begin{align}
X_0(\xi) &= \frac{1}{w}\quad (w \equiv 1-b\xi)\\
X_1(\xi) &= \frac{-c\ln w}{w^2}\\
X_2(\xi) &= \frac{c^2(\ln^2 w - \ln w+w-1)-c_2(w-1)}{w^3}
\end{align}
\end{subequations}
\begin{subequations}
\begin{align}
Y_0(\xi) &= w^\rho \quad \\
Y_1(\xi) &= \rho w^{\rho-1} \quad \left[ c(\ln w) + (1-w)(c-g_1)\right]\\
Y_2(\xi) &= \frac{\rho}{2} w^{-2+\rho} \big\{ -c^2(1-\rho)\ln^2w + \left[ -2\rho(c-g_1)(w-1)+2g_1\right]c\ln w\nonumber  \\
 &+ (w-1) \left[ \left(c^2 - c_2 +\rho (c-g_1)^2\right)(w-1) + (g_2 - cg_1)(w+1)\right] \big\}.
\end{align}
\end{subequations}
We note that $X_n$ depends on $c_m(c_0 \equiv b, c_1 \equiv c)$ for $m = 0,1 \ldots n$ and $Y_n$ depends on $c_m$ and $g_m (g_0 \equiv f)$ for $m = 0,1 \ldots n$.  In general, $c_m$ and $g_m$ are computed through evaluation of Green's functions with $1,2\ldots(m+1)$ loops. 

The mass parameter $I\!\!M$ was introduced in eq. (13).  Written in the form
\begin{equation}
I\!\!M = \frac{m(\mu)}{E(a(\mu))}
\end{equation}
it appears to be a RS dependent quantity.  However, by direct differentiation with respect to each of the RS parameters $c_i$ and $g_i$, we show that the exact expression for $I\!\!M$ is, in fact, RS independent.
\begin{subequations}
\begin{align}
\frac{\partial I\!\!M}{\partial c_i} &= m(\mu) \left[ \Gamma_i^c (a) - B_i(a) \frac{\gamma(a)}{\beta(a)} - \int_0^a dx \frac{bx^{i+2}\gamma(x)}{\beta^2(x)}\right] E^{-1}(a(\mu))= 0\\
\frac{\partial I\!\!M}{\partial g_i} &= m(\mu) \left[ \Gamma_i^g (a) -  \int_0^a dx   \frac{fx^{i+1}}{\beta(x)}\right] E^{-1}(a(\mu))= 0.\\
\intertext{By a similar computation, it also follows that}
\mu \frac{\partial I\!\!M}{\partial \mu} &= 0.
\end{align}
\end{subequations}
It should be noted however that even though an exact expression for $I\!\!M$ given by eq. (31) is independent of $\mu$, $c_i$ and $g_i$, any approximation for $I\!\!M$ based on a finite number of terms contributing to $m(\mu)$ or $E(a(\mu))$ will have some dependence on these parameters.

The full resummed expression of eq. (23) for the decay rate $\Gamma$ is RS independent.  However, $a\left( \ln \frac{I\!\!M}{\Lambda}\right)$ and  $m \left( \ln \frac{I\!\!M}{\Lambda}\right)$ both display RS dependence through the $c_i$ and $g_i$ according to eq. (27).  It follows that the expansion coefficients $T_n \equiv T_{n,0}$ must also be explicitly RS dependent.  This RS dependence can be deduced from
\begin{subequations}
\begin{align}
\frac{\partial \Gamma}{\partial c_i} &= 0 = m^5 \sum_{n=0}^\infty \left[ 5\Gamma_i^c(a)T_n a^n + \frac{\partial T_n}{\partial c_i} a^n + n B_i(a) T_n a^{n-1}\right] \quad (i \geq 2)\\
\frac{\partial \Gamma}{\partial g_i} &= 0 = m^5 \sum_{n=0}^\infty \left[ 5\Gamma_i^g(a)T_n a^n + \frac{\partial T_n}{\partial g_i} a^n 
\right] \quad (i \geq 1)
\end{align}
\end{subequations}
where $a \equiv a\left( \ln \frac{I\!\!M}{\Lambda}\right)$ and $m \equiv m\left( \ln \frac{I\!\!M}{\Lambda}\right)$, by using the
explicit form of the $W_n^i$, $U_n^i$ and $V_n^i$ expansion coefficients referred to earlier.  In particular, we have $U^2_0=-\rho/2$, $V^2_0=-\rho/2$, $V^1_0=-\rho$ and $V^1_1=c\rho/2$. For $n = 0,1,2$ we find that 
\begin{subequations}
\begin{align}
&\frac{\partial T_0}{\partial c_i} = 0\\
&\frac{\partial T_1}{\partial c_i} = 0\\
&\frac{\partial T_2}{\partial c_i} + 5\delta_2^i\left(T_0 U_0^i\right) = 0
\end{align}
\end{subequations}
\begin{subequations}
\begin{align}
&\frac{\partial T_0}{\partial g_i} = 0\\
&\frac{\partial T_1}{\partial g_i} + 5\delta_1^i \left(T_0 V_0^i\right) = 0\\
&\frac{\partial T_2}{\partial g_i}  + 5\delta_2^i\left( T_0 V_0^i\right) + 5\delta_1^i \left(V_0^1T_1 + V_1^1T_0\right) = 0.
\end{align}
\end{subequations}
Solving these equations leads to
\begin{subequations}
\begin{align}
& T_0 = \tau_0\\
&T_1 = \tau_1 + 5\rho \tau_0 g_1\\
&T_2 = \tau_2 + \frac{5\rho}{2}\left(\tau_0c_2 +\tau_0 g_2 -c \tau_0 g_1 +2 \tau_1g_1 + 5\rho\tau_0g_1^2\right)  
\end{align}
\end{subequations}
where $\tau_0$, $\tau_1$, $\tau_2$ are constants of integration and, consequently, RS independent.  An iterative analysis of eq. (33) will generate the explicit solutions for $T_n (n \geq 3)$ in terms of the $c_i$ and $g_i (i = 0,1 \ldots n)$ expansion coefficients and further RS independent variables $\tau_n (n \geq 3)$. The $\tau_i$ are related to the RS invariants derived in ref. [25].

Since $\tau_0, \tau_1 \ldots$ are RS independent, their values can be identified in any RS from a perturbative calculation of the decay rate $\Gamma$ in eq. (23).  If an explicit evaluation of $T_n$, $c_i$ and $g_i$ has been carried out to N$^{th}$ order in perturbation theory through the evaluation of Feynman diagrams using some mass-independent RS, such as $\overline{MS}$ [2,15], then $\tau_0, \tau_1 \ldots \tau_N$ can be deduced for this process by availing eqs. (36) above and their higher order iterative companion equations.  It is interesting that the decay rate expansion coefficient $T_n$, in general, is computed by a calculation involving $n$-loop Green’s functions.  However, the identification of the value of $\tau_n$ involves $c_i$, $g_i (i = 0,1 \ldots n)$ where $c_n$ and $g_n$ are obtained by considering $n + 1$ loop Green's functions.  Furthermore, identification of the decay rate expansion coefficient $T_n$, under a change of RS, involves alterations dependent on the same set of parameters, of which $c_n$ and $g_n$ are again obtained by considering $(n+1)$-loop Green’s functions.

\section{The Pole Mass and the Running Mass}

In eq. (23) we have an expression for $\Gamma$ that depends on a mass scale $I\!\!M$ which is essentially a boundary value for the equation for the running mass $m(\mu)$ of eq. (5).  We will now relate this mass scale $I\!\!M$ to the pole mass $m_{\rm{pole}}$ of the $b$ quark.  This pole mass is a RS independent, gauge invariant and infrared finite quantity [16].  Since quarks are always in a bound state, one cannot directly measure this pole mass; it is a quantity that is realized in perturbation theory.  In a number of papers the self energy of the quark is discussed in detail and from this the relationship between $m_{\rm{pole}}$ and $m(\mu)$ can be derived [17,35] (see also ref. [36]).  From this, a relation can be derived between $I\!\!M$ and $m_{\rm{pole}}$.

If one uses a mass independent RS, the inverse of the renormalized quark propagator has the form
\begin{equation}
S^{-1}\left(p^\mu, m(\mu)\right) = A \left(p^2, m(\mu)\right)\diagup{\!\!\!\!p} - m(\mu) 
B\left(p^2, m(\mu)\right);
\end{equation}
the pole mass is to be the location of the zero of this inverse propagator
\begin{equation}
\lim_{\diagup{\!\!\!\!p} \rightarrow m_{\rm{pole}}}\quad S^{-1} \left(p^\mu, m(\mu)\right) = 0.
\end{equation}
This results in an expansion
\begin{equation}
 m_{\rm{pole}} = m(\mu) \sum_{n=0}^\infty \sum_{k=0}^n S_{n,k} a(\mu)^n L^k
\end{equation}
where $S_{0,0} = 1$ and $L = \ln\left( \frac{\mu}{m_{\rm{pole}}}\right)$.  The coefficients $S_{n,k}$ in eq. (39) are also known numerically to four loop order [35].  In ref. [18] it is pointed out that for the $b$ quark, this expansion for the pole mass indicates that the relation between $m_{\rm{pole}}$ and $m(\mu)$ is likely to be affected by divergences due to renormalons, beginning at third order, and that it might be more appropriate to use $m(\mu)$ rather than $m_{\rm{pole}}$ when discussing the $b$ quark.

The structure of the equation relating $m_{\rm{pole}}$ to $m(\mu)$ is very similar to that of eq. (1) for the decay width $\Gamma$.  We can now reorganize the summations in eq. (39) in a similar manner to the analysis of section two.

We define (as in eq. (7))
\begin{equation}
F_n(a(\mu)) = \sum_{k=0}^\infty S_{n+k,n} a(\mu)^{n+k}
\end{equation}
so that 
\begin{equation}
m_{\rm{pole}} = m(\mu) \sum_{n=0}^\infty F_n (a(\mu))L^n.
\end{equation}
We know that $m_{\rm{pole}}$ is $\mu$ independent so that 
\begin{equation}
\mu \frac{dm_{\rm{pole}}}{d\mu} = 0 = \left( \mu \frac{\partial}{\partial\mu} + \beta(a) \frac{\partial}{\partial a} + m\gamma(a) \frac{\partial}{\partial m}\right)\left[ m \sum_{n=0}^\infty F_n(a)L^n\right].
\end{equation}
This can be used to show that
\begin{equation}
F_{n+1}(a) = - \frac{1}{n+1} \left( \beta(a) \frac{\partial}{\partial a} + \gamma(a)\right)F_n(a).
\end{equation}
With $E$ defined in eq. (11) and
\begin{equation}
\phi_n = EF_n
\end{equation}
then by eq. (43)
\begin{equation}
\phi_{n+1}(a) = - \frac{1}{n+1} \beta(a) \frac{d}{da} \phi_n(a)
\end{equation}
or
\begin{equation}
\phi_{n+1} \left( a\left( \ln \frac{\mu}{\Lambda}\right)\right) = -\frac{1}{n+1} 
\frac{d}{d\left( \ln \frac{\mu}{\Lambda}\right)} \phi_n \left( a\left( \ln \frac{\mu}{\Lambda}\right)\right).
\end{equation}
We thus find that
\begin{equation}
m_{\rm{pole}} = m(\mu)E^{-1}\left( a(\mu)\right) \sum_{n=0}^\infty \frac{(-L)^n}{n!} \frac{d^n}{d \ln\left(\frac{\mu}{\Lambda}\right)^n} \phi_0\left(a\left( \ln \frac{\mu}{\Lambda}\right)\right)
\end{equation}
\begin{equation}
= I\!\!M \phi_0 \left( a\left( \ln \frac{\mu}{\Lambda} - L\right)\right),
\end{equation}
or by eqs. (14) and (13)
\begin{subequations}
\begin{align}
m_{\rm{pole}} &= I\!\!M E\left( a \left( \ln \frac{m_{\rm{pole}}}{\Lambda}\right)\right) F_0\left( a\left( \ln\frac{m_{\rm{pole}}}{\Lambda}\right)\right) \\
&= m\left( \ln \frac{m_{\rm{pole}}}{\Lambda}\right) F_0\left( a\left( \ln\frac{m_{\rm{pole}}}{\Lambda} \right)\right).
\end{align}
\end{subequations}

The resummations carried out on the all-orders expansion of eq. (39) lead to eq. (49b) in which the b-quark pole mass is related to the running b-quark mass evaluated at the pole mass and the all orders series
\begin{equation}
F_0\left(a\left (\ln \frac{m_{pole}}{\Lambda} \right)\right) = \sum_{n=0}^\infty S_{n,0} a^{n}\left(\ln \frac{m_{pole}}{\Lambda}\right)
\end{equation}

As a by-product of this analysis, we find a second approach to the identification of the mass scale $I\!\!M$ for use in the decay rate expansion of eq. (23b) and via eqs. (49a,b) to obtain
\begin{equation}
I\!\!M = \frac{m_{\rm{pole}}}{ E\left(a \left( \ln \frac{m_{\rm{pole}}}{\Lambda}\right)\right) F_0\left( a\left( \ln\frac{m_{\rm{pole}}}{\Lambda}\right)\right)}
\end{equation}

Availing eq. (51) for $I\!\!M$ allows us to find the decay rate $\Gamma$ in terms of the pole mass and the running coupling evaluated at the pole mass.

It is now possible to examine the RS dependency of the expansion coefficients $S_n \equiv S_{n,0}$ of the function $F_0$ defined in eq. (40). We find that since $m_{\rm{pole}}$ in eq. (49) is RS invariant, then
\begin{equation}\tag{52a-c}
\mu \frac{dm_{\rm{pole}}}{d\mu} = 0 =  \frac{dm_{\rm{pole}}}{d c_i} = 
\frac{dm_{\rm{pole}}}{d g_i} . 
\end{equation}
Using eqs. (40,49) in eqs. (52b,c) leads to 
\begin{align}
m \sum_{n=0}^\infty &\left[ \Gamma_i^c(a) S_n a^n + \frac{\partial S_n}{\partial c_i} a^n + n B_i(a) S_n a^{n-1}\right] = 0, i \geq 2 \tag{53a} \\
m \sum_{n=0}^\infty &\left[ \Gamma_i^g(a) S_n a^n + \frac{\partial S_n}{\partial g_i} a^n\right] = 0, (i \geq 1) \tag{53b}
\end{align}
where $a = a\left( \ln \frac{m_{\rm{pole}}}{\Lambda}\right)$ and $m = m\left( \ln \frac{m_{\rm{pole}}}{\Lambda}\right)$.  For $n = 0,1,2$ we find 
\begin{align}
\frac{\partial S_0}{\partial c_i} &= 0, \quad \frac{\partial S_1}{\partial c_i} = 0, \quad \frac{\partial S_2}{\partial c_i} - \frac{\rho}{2} S_0\delta_2^i = 0 \tag{54a-c}\\
\intertext{and}
\frac{\partial S_0}{\partial g_i} &= 0, \quad \frac{\partial S_1}{\partial g_i} - \rho S_0\delta_1^i = 0, \quad \frac{\partial S_2}{\partial g_i}  + \left( - \rho S_1 + \frac{\rho c}{2} S_0\right) \delta_1^i - \frac{\rho}{2} S_0 \delta_2^i = 0.
\tag{55a-c}
\end{align}
Solving these equations leads to 
\begin{equation}\tag{56a}
S_0 = \sigma_0
\end{equation}
\begin{equation}\tag{56b}
S_1 = \sigma_1 + \rho \sigma_0 g_1^2
\end{equation}
\begin{equation}\tag{56c}
S_2 = \sigma_2 +\frac{\rho}{2}\left( \sigma_0 c_2 + 2\sigma_1 g_1- c \sigma_0 g_1+ 2\rho g_2 g_1\sigma_0 + \sigma_0 g_2 \right)
\end{equation}
In eq. (56), the quantities $\sigma_i$ are constants of integration and thus are RS invariants.  Similar expressions can be found for all $S_n(n \geq 0)$. It is again interesting to note the relationship between expansion coefficients (in this case, those of the pole mass $S_n(n \geq 0)$ and RS invariant parameters $\sigma_n(n>0)$). The $S_n$ coefficients are found by computing n-loop Feynman integrals. However, the identification of the values of $\sigma_n (n>0)$ involves $c_i$ and $g_i$ $(i = 0,1,…,n)$ where $c_n$ and $g_n$ are obtained by consideration of $(n+1)$-loop Feynman integrals. Furthermore, using eqs. (56a-c) and their iterative companion equations to determine the $n$-loop expansion coefficients $S_n$ for the pole mass (when relating it to the running mass and running coupling in a different RS), we see that the  alterations of the $S_n$ involve changing the parameters $c_n$ and $g_n$ which are obtained from $(n+1)$-loop Feynman integrals. This dependence of the $n$-loop quantity $S_n$ on $(n+1)$-loop parameters $(c_n,g_n)$ is much like the dependence of $T_n$ on $(c_n,g_n)$ noted after eq. (36). 

If it were possible to make an ansatz for $S_n$ on the parameters $c_i$ and $g_i$ that characterizes the RS, then summation such as the one that leads to the $\mu$-independent result of eq. (49b) would be feasible and all dependence on the RS parameters $c_i$ and $g_i$ would cancel. It is quite possible that the RS independent result would be identical to what would be obtained if we were to get $c_i=g_i=0$ at the outset so that $S_n=\sigma_n$, the RS independent parameters arising in eq. (56).  
 
\section{Evaluation of $\Gamma$}

In order to illustrate how the formalism of the preceding sections can be applied, we now use it to compute $\Gamma$ in perturbation theory within the $\overline{MS}$ RS.  Two other interesting RS will also be considered.

From ref. [7,18] we know that the coefficients $T_n \equiv T_{n,0}$ appearing in eq. (1) are for $n = 0,1,2$ (for $SU(3)$ with five flavours)
\begin{align}
T_0 &=1 \tag{57a}\\
T_1 &= \frac{65}{6} - \frac{2\pi^2}{3}\tag{57b}\\
T_2 = \frac{3763495}{23328} - \frac{38\pi^2\ln 2}{27} &- \frac{35453\pi^2}{2916} - \frac{3259\zeta (3)}{81} + \frac{289\pi^4}{648}\tag{57c}
\end{align}
when using $\overline{MS}$.   

The known exact $\overline{MS}$ values up to two-loop are now used to find the first few terms in the perturbative expansion for $A_0\left(a\left(\ln\frac{I\!\!M}{\Lambda}\right)\right)$
\begin{equation}
A_0 \approx T_0 a+ T_1a^2 + T_2a^3\nonumber \\
\end{equation}
appearing in eq. (23). (Via Pad\'e\ approximation, $T_3$ can be estimated [19,20].) To find $\Gamma$ itself in eq. (23), we need first to determine $I\!\!M$.  By eq. (32) we know that the exact expression for $I\!\!M$ is independent of $\mu$, $c_i$ and $g_i$, so that it is RS independent, and consequently we estimate $I\!\!M$ by using perturbative expansions for $m(\mu)$ and $E(a(\mu))$ in the $\overline{MS}$ RS. Since we are using perturbative expansions to find $I\!\!M$, what we will use will retain dependence on $\mu$ and the RS used, even though the exact expression for $I\!\!M$ is $\mu$ and RS independent. 

From eq. (11), as $K \rightarrow \infty$, the two loop estimate for $E$ is given by
\begin{align}
E^{(2)} (a) &= \exp \left[ \int_0^a dx \frac{fx(1+g_1x)}{-bx^2(1+cx)} + \int_0^\infty dx \frac{fx}{bx^2(1+cx)}\right]\nonumber \\
&= \left[ \frac{(1+ca)^{-1+g_1/c}}{ca}\right]^{f/b}\tag{58}
\end{align}
while at three loop order, we also provide an analytic expression, 
\begin{equation}
E^{(3)}(a) = \exp \left[ \int_0^a dx \frac{f(1+g_1 x + g_2x^2)}{-bx^2(1+cx + c _2x^2)} + \int_0^\infty dx \frac{fx}{bx^2(1+cx)}\right]\nonumber
\end{equation}
which for $4c_2 - c^2 > 0$ (which is satisfied in our situation) leads to
\begin{align}
\ln E^{(3)} (a) = &- \frac{f}{b}\ln (ca) + \frac{f}{2b} \left(1 - \frac{g_2}{c_2}\right) \ln (1 + ca + c_2a^2)\tag{59}\\
&- \frac{f}{b} \left(g_1 - \frac{c}{2}\left(1 + \frac{g_2}{c_2}\right)\right) \frac{2}{\sqrt{4c_2-c^2}} \Big[ \tan^{-1} \left( \frac{c+2c_2a}{\sqrt{4c_2-c^2}}\right)\nonumber \\
&- \tan^{-1} \left(\frac{c}{\sqrt{4c_2-c^2}}\right)\Big].\nonumber
\end{align}
In the limit $g_2 \rightarrow 0$ followed by the limit $c_2 \rightarrow 0$, we find that $E^{(3)} (a) \rightarrow E^{(2)}(a)$. 

The mass of the $b$ quark is generally presented in one of two ways.  One way is in terms of the ``$\overline{MS}$ mass'' (which we will denote by $\overline{m}_b$).  In this approach, $\overline{m}_b$ is used to fix the boundary condition on the running mass so that
\begin{equation}
\overline{m}_b = m (\overline{m}_b),\tag{60}
\end{equation}
and so by eq. (31) we have upon choosing $\mu = \overline{m}_b$
\begin{equation}
I\!\!M = \frac{\overline{m}_b}{E(a(\overline{m}_b))}.\tag{61}
\end{equation}
The generally accepted value of $\overline{m}_b$ is [21]
\begin{equation}
\overline{m}_b = 4.18 \ GeV\nonumber
\end{equation}
and also for the $SU(3)$ colour gauge group with five flavours using the $\overline{MS}$ RS
\begin{equation}
b = \frac{23}{6}, \quad c = \frac{29}{23}, \quad c_2 = 1.474789 \tag{62a-c}
\end{equation}
\begin{equation}
f = -2, \quad g_1 = \frac{253}{72}, \quad g_2 = 3.70993. \tag{63a-c}
\end{equation}
In refs. [32,33], $c_2$ was first calculated while $g_2$ appeared first in ref. [34]. The value of the running coupling $a(\mu)$ is generally given at the $Z$ boson mass [21]
\begin{equation}
\mu = I\!\!M_z = 91.1876 \ GeV \tag{64}
\end{equation}
to be
\begin{equation}
a\left(\ln \frac{I\!\!M_z}{\Lambda}\right) = \frac{.1185}{\pi}=0.03772.\tag{65}
\end{equation}

Using the initial conditions above, we demonstrate the scale dependence of the running $\overline{MS}$ coupling $a$ and b-quark mass, $m_b$ using eqs. (28,29,30)  in Figures 1 and 2 respectively. Based on the 2- and 3-loop results derived above, we also plot the $\mu$ dependence of $I\!\!M$ of eq. (31) in Figure 3. Here, we find that the 3-loop expression is relatively more scale independent than the 2-loop expression for $I\!\!M$ as expected, since the exact expression is scale-independent. 

To obtain $a\left(\ln \frac{\overline{m}_b}{\Lambda}\right)$, needed for $I\!\!M$ in eq. (61), we use eqs. (29a), (29b) for $X_0$ and $X_1$ in eq. (28a) with
\begin{equation}
a\left(\ln \frac{\overline{m}_b}{\Lambda}\right) \approx \sum_{n=0}^1 X_n\left( a\left(\ln \frac{I\!\!M_z}{\Lambda}\right) \ln 
\left( \frac{I\!\!M_z}{\overline{m}_b}\right)\right) a^n \left(\ln \frac{I\!\!M_z}{\Lambda}\right).\tag{66}
\end{equation}
This leads to
\begin{equation}
a\left(\ln \frac{\overline{m}_b}{\Lambda}\right) = .07150\tag{67}
\end{equation}
so that in eq. (61) with $E\left( a\left(\ln \frac{\overline{m}_b}{\Lambda}\right)\right)$ being approximated by 
$E^{(2)}\left(a\left(\ln \frac{\overline{m}_b}{\Lambda}\right)\right)$ we have 
\begin{equation}
I\!\!M = 15.90 \ GeV .\tag{68}
\end{equation}

In obtaining this value of $I\!\!M$, we have used $\mu = \overline{m}_b$ in eq. (31).  In place of this, we could choose $\mu = I\!\!M_z$ with
\begin{equation}
m \left(\ln \frac{I\!\!M_z}{\Lambda}\right) \approx \overline{m}_b \sum_{n=0}^1 Y_n \left( a\left(\ln \frac{\overline{m}_b}{\Lambda}\right)
\ln \frac{\overline{m}_b}{I\!\!M_z}\right)a^n\left(\ln \frac{\overline{m}_b}{\Lambda}\right)\tag{69}
\end{equation}
with $Y_0$ and $Y_1$ being given in eq. (30).  We could also invert eq. (69) so that
\begin{equation}
m \left(\ln \frac{I\!\!M_z}{\Lambda}\right) \approx \overline{m}_b \left[ \sum_{n=0}^1 Y_n
 \left( a\left(\ln \frac{I\!\!M_z}{\Lambda}\right)
\ln \frac{I\!\!M_z}{\overline{m}_b}\right)a^n\left(\ln \frac{I\!\!M_z}{\Lambda}\right)\right]^{-1}.\tag{70}
\end{equation}
From this equation for $m\left(\ln\frac{I\!\!M_z}{\Lambda}\right)$ and with $E^{(2)}\left( a\left(\ln\frac{I\!\!M_z}{\Lambda}\right)\right)$, we find from eq. (31) that
\begin{equation}
I\!\!M \approx 14.70 \ GeV \tag{71}
\end{equation}

The values of $I\!\!M$ obtained to second order in perturbation theory in eqs. (68) and (71) using respectively $\mu = \overline{m}_b$ and $\mu = I\!\!M_z$ are consistent with the exact expression for $I\!\!M$ being independent of $\mu$ (see. eq. (32)).

We now can compute the perturbative result for $\Gamma$ given by eq. (23) using $T_0$, $T_1$ and $T_2$ given in eq. (57).  We find that our perturbative result for $\Gamma$ is 
\begin{equation}
\Gamma \approx \left[ m\left(\ln \frac{I\!\!M}{\Lambda}\right)\right]^5\left[T_0 + T_1 a\left(\ln \frac{I\!\!M}{\Lambda}\right) + T_2 
a\left(\ln \frac{I\!\!M}{\Lambda}\right)^2\right].\tag{72}
\end{equation}
In this equation $m\left(\ln \frac{I\!\!M}{\Lambda}\right)$ and $a\left(\ln \frac{I\!\!M}{\Lambda}\right)$ for $I\!\!M$ given by eqs. (68) and (71), we first use eq. (28) up to $n = 1$ (i.e. two-loop order) with $\mu = I\!\!M$ and $\mu^\prime = \overline{m}_b$ and $I\!\!M_z$ respectively.  We find that for these two values of $I\!\!M$, the corresponding values of $\Gamma$ coming from eq. (72) are
\begin{align}
\Gamma &= 615.2 \ GeV^5\tag{73}\\
\intertext{and}
\Gamma &= 648.8 \ GeV^5\tag{74}
\end{align}
respectively. When $\mu = \overline{m}_b$ (as in the case of eq. (73)), we see by eqs. (6,60) that the two loop result for $\Gamma$ corresponds to the RG summed result of eq. (23) but with  $I\!\!M$ replaced by $\overline{m}_b$. These two values of $\Gamma$ differ by 2.66$\%$ from the mid point. In Figure 4, we plot the renormalization scale dependence of the 2-loop perturbative expression as compared to the scale invariant results provided in Eq. (73) and Eq. (74). We unexpectedly find that the Eq. (73) value (where $\mu = \overline{m}_b$ in computing $I\!\!M$) intersects with the perturbative curve at $\mu = I\!\!M$, while the Eq. (74)  (where $\mu = I\!\!M_z $ in computing $I\!\!M$) value intersects at $\mu = 7.6 \ GeV$. This key finding shows how our scale invariant expression indicates that a choice of $\mu = I\!\!M$ renders it equivalent to the two-loop perturbative expression, which implies as to how this renders a choice of $\mu=\overline{m}_b$ scale dependent, even though this latter choice sets the RG log terms to zero in the perturbative expression. Overall, our approach eliminates errors in $\Gamma$ due to scale dependence. These errors are typically ascertained using the perturbative expression by variation of the decay rate in the range of $\overline{m}_b/2\leq\mu\leq2\overline{m}_b$.  

If one were to use three loop contributions to $\beta$ and $\gamma$ but expand $\Gamma$ only to second order in $a$ (since the third order coefficient $T_3$ is unknown), then the values of $\Gamma$ corresponding to $\mu = \overline{m}_b$ and $\mu = I\!\!M_z$ are
\begin{align}
\Gamma &= 672.4 \ GeV^5 \tag{75}\\
\intertext{and}
\Gamma &= 677.8 \ GeV^5 \tag{76}
\end{align}
respectively.   These values differ by 0.41$\%$ about the mid point.  We thus see that the uncertainty in $\Gamma$ is significantly decreased by going to higher order in perturbation theory.

Even though $\Gamma$ is a physical quantity and is consequently RS independent, any perturbative approximation to $\Gamma$ is RS dependent.  One generally uses the $\overline{MS}$ RS when doing an explicit calculation as it is the easiest RS to employ, but it should be kept in mind that $c_i$ and $g_i$ are RS dependent with the dependency of $a$, $m$ and $T_n$ on these parameters fixed by eqs. (26,34) respectively.  In refs. [8,9], it is suggested that the optimal values of $c_i$ and $g_i$ that should be used in any perturbative result is determined by the principle of minimal sensitivity (PMS).  This means that the optimal values of these parameters that characterize a RS are those whose variation least affects the value of $\Gamma$.

The form of the explicit dependence of $\Gamma$ on $\mu$ appearing in eq. (1) is essential for being able to make use of the RG equation to arrive at the $\mu$-independent result of eq. (23).  This form comes from knowing the way in which Feynman diagrams give rise to dependence  of $\Gamma$ on $\mu$ at $n$-loop order.   Unfortunately, there does not appear to be any way of arriving at a way of seeing in general how $T_{n,k}$ in eq. (1) depends on $c_i$, $g_i$ and consequently a summation analogous to that of eq. (7) is not possible.  If such a summation were feasible, then the dependence of $\Gamma$ on $c_i$ and $g_i$ would drop out in a way analogous to the way the dependence of $\Gamma$ on $\mu$ has disappeared in eq. (23).

There are some RS in addition to $\overline{MS}$ that are of particular interest.  In the first scheme, the 't Hooft scheme [11], one simply sets $c_i = g_i = 0$.  In this case, we see from eqs. (36,56) that $T_n = \tau_n$ and $S_n = \sigma_n$.  What makes this scheme interesting is that the perturbative series for $\beta$ and $\gamma$ truncates so that
\begin{align}
\mu \frac{da}{d\mu} &= -ba^2(1 + ca)\tag{77}\\
\mu \frac{dm}{d\mu} &= m fa \Rightarrow \frac{dm}{da} = - \frac{f}{b}\frac{m}{a(1+ca)},\tag{78}
\end{align}
making it possible to find $a(\mu)$ and $m(a(\mu))$.  (The Lambert $W$ function occurs in $a(\mu)$ in this RS [22]).

In a second scheme, we choose $c_i$ and $g_i$ so that $T_n = 0$ in the series for $A_0$ for $n \geq 1$ in eq. (7).  By eq. (36b), this means that if $T_1 = 0$, then
\begin{equation}\tag{79a}
g_1 = \frac{-\tau_1}{5\rho\tau_0}
\end{equation}
while if $T_2 = 0$, then (36)
\begin{equation}\tag{79b}
c_2 + g_2 = -\frac{2}{5\rho \tau_0}  \left[ \tau_2 + \frac{5\rho g_1}{2}(-c\tau_0+2\tau_1+5\rho\tau_0g_1)\right] 
\end{equation}
etc.  Having the values of $c_i$ and $g_i$ fixed in this manner alters the functions $\beta$ and $\gamma$ but reduces $\Gamma$ to the single contribution $T_0$ appearing in $A_0(a)$.

A related scheme is to choose $c_i$ and $g_i$ so that the series for $F_0$ in eq. (40) terminates at $S_0$.  By eq. (56) this means that
\begin{equation}\tag{80a}
g_1 = \frac{-\sigma_1}{\rho\sigma_0}
\end{equation}
for $S_1 = 0$,
\begin{equation}\tag{80b}
c_2 + g_2 = -\frac{2}{\rho \sigma_0}\left( \sigma_2 +\frac{\rho g_1}{2}(2\sigma_1-c\sigma_0+\rho\sigma_0g_1)\right)
\end{equation}
for $S_2 = 0$, etc.  Again $\beta$ and $\gamma$ are altered with these conditions, but now the series for $F_0$ is reduced to a single term $S_0$.

\section{Discussion}

It is well known that a perturbative evaluation to finite order of radiative corrections to a physical process $\Gamma$ has dependence on the renormalization mass scale $\mu$ and the RS parameters ($c_i, g_i$), all of which are unphysical.  However, the way in which one computes Feynman diagrams makes it possible to determine the explicit form of the dependence of $\Gamma$ on $\mu$ at each order of perturbation theory (see. eq. (1)).  The RG equation is derived by requiring that this explicit dependence of $\Gamma$ on $\mu$ must be compensated by an implicit dependence of $\Gamma$ on $\mu$ through a ``running coupling'' $a(\mu)$ and a ``running mass'' $m(\mu)$.  We have shown how this RG equation makes it possible to sum all radiative contributions to $\Gamma$ that depend explicitly on $\mu$ and that upon doing this, the explicit and implicit dependence of $\Gamma$ on $\mu$ cancels so that in place of $\mu$, $\Gamma$ depends on two mass scales $\Lambda$ and $I\!\!M$ which are associated with boundary conditions on the running parameters $a(\mu)$ and $m(\mu)$ respectively.  In practice, one does not directly deal with $\Lambda$ and $I\!\!M$, but rather with $a(\mu)$ and $m(\mu)$ at an experimentally determined mass scale.  For $a(\mu)$, $\mu$ is generally taken to be $I\!\!M_z$, while for $m(\mu)$ we use $\mu = \overline{m}_b$ which is the ``$\overline{MS}$ mass'' at which $\overline{m}_b = m(\overline{m}_b)$.  It is also possible to use $m_{\rm{pole}}$, the ``pole mass'' defined in eq. (28), as the mass scale.  This pole mass for the $b$ quark is $m_{\rm{pole}} = 4.7659 GeV$ [21]; we could use $a(m_{\rm{pole}})$, $m(m_{\rm{pole}})$ to compute $\Gamma$.

Using these boundary values for $a(\mu)$ and $m(\mu)$, we have been able to deduce $I\!\!M$.  This involves determining how $a(\mu)$ and $m(\mu)$ vary under changes of $\mu$ using a summation of leading-log, next-to-leading-log etc. contributions to variation of $a(\mu)$; this has been carried out using up to the three-loop contributions to $\beta$ and $\gamma$.  This virtually eliminates dependency of 
$I\!\!M$ on $\mu$, though there remains RS dependency residing in $c_i$ and $g_i$.  In addition $\Gamma$ depends on a set of RS independent parameters $\tau_i$ and $\sigma_i$.  One can, in principle, choose $c_i$ and $g_i$ to reduce $\beta$ and $\gamma$ to a finite series in $a$, or alternatively, these parameters could be chosen to reduce the expansion of $\Gamma$ in eq. (23) to a finite number of terms.  It is not clear how the latter choice of RS would affect the nature of the perturbative expansion of $\Gamma$ (in particular how ``renormalons'' occur [10,13].)  It has been proposed [8,9] that the optimal values of $c_i$ and $g_i$ used to compute $\Gamma$ are those chosen so that $\Gamma$ is stable under small variations away from these values.  This is the ``principle of minimal sensitivity''.  This involves satisfying the conditions
\begin{equation}\tag{81a,b}
\frac{\partial\Gamma}{\partial c_i} = 0 = \frac{\partial\Gamma}{\partial g_i}
\end{equation}
at each order of perturbation theory.  We have not attempted in this paper to select $c_i$ and $g_i$ using these conditions but rather have just worked in the $\overline{MS}$ RS. In a future study [24], we intend to conduct a systematic error analysis of $|V_{ub}|$ with considerations of non-perturbative contributions and systematic estimation of scheme dependence errors. We have obtained higher order expressions at 3-loop order and would also draw from estimates [19,20] in order to conduct this analysis.  

It would be interesting if there were a way of explicitly summing all of the explicit dependence of $\Gamma$ on $c_i$ and $g_i$ so as to have a cancellation of the explicit and implicit dependence of $\Gamma$ on these parameters, in much the same way that all dependence on $\mu$ has been shown above to cancel.  One might expect the final result of such a summation would be the same as what is obtained in the 
't Hooft RS in which $c_i = g_i = 0$, so that $T_n = \tau_0$ and $S_n = \sigma_n$. Moreover, an extensiom of this type of analysis to the situation in which there is more than one coupling [23] or mass present would be interesting.

\section*{Acknowledgements}
We would like to thank A. Kataev for several very helpful suggestions.  R. Macleod initiated this work.

\newpage

\begin{figure}[hbt]
\begin{center}
\includegraphics[scale=0.36]{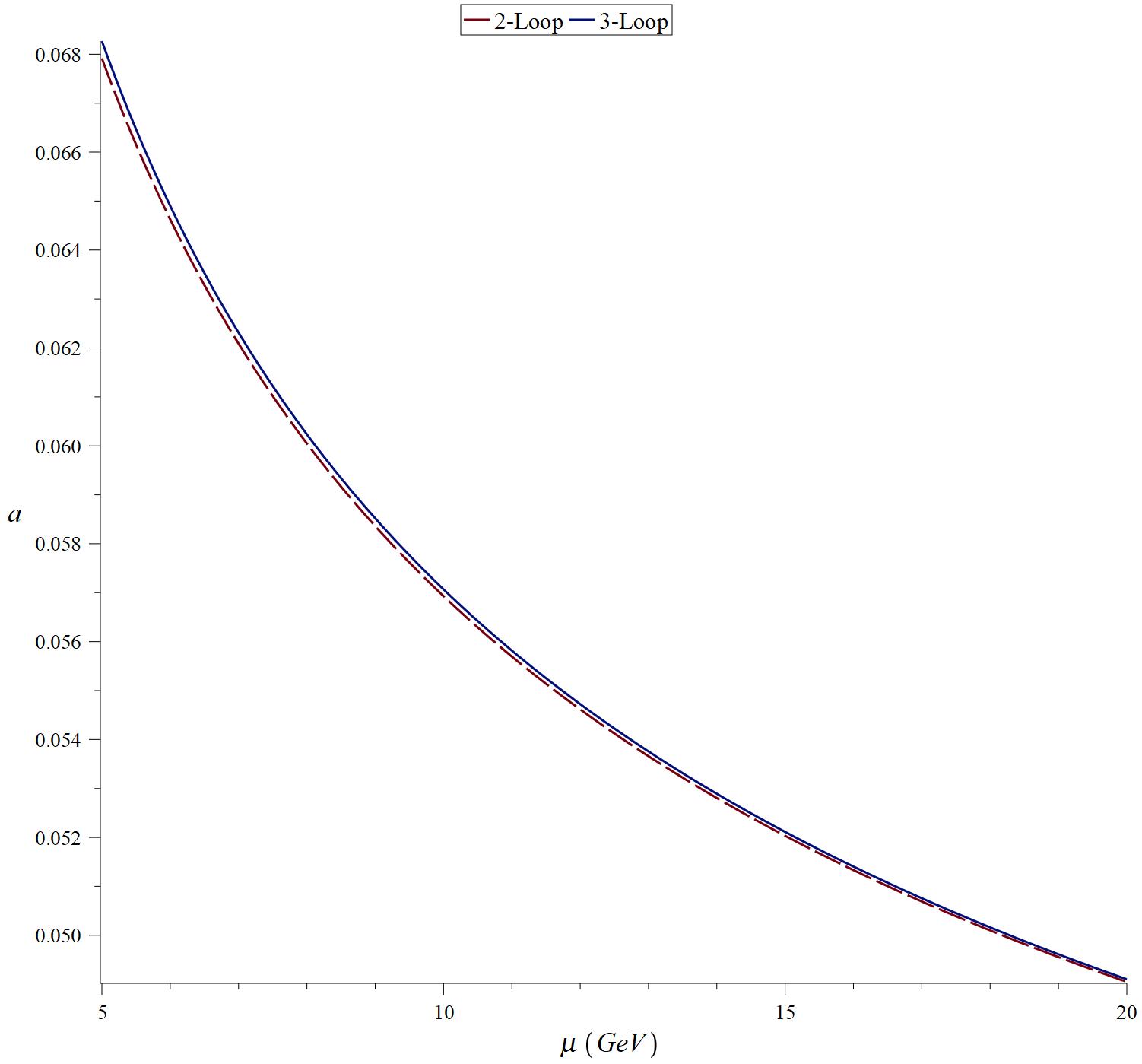}
\caption{The renormalization scale dependence of strong coupling $a$ in the ${\overline{MS}}$ scheme at 2- and 3-loop orders}
\label{Fig. 1}
\end{center}
\end{figure}

\newpage

\begin{figure}[hbt]
\begin{center}
\includegraphics[scale=0.36]{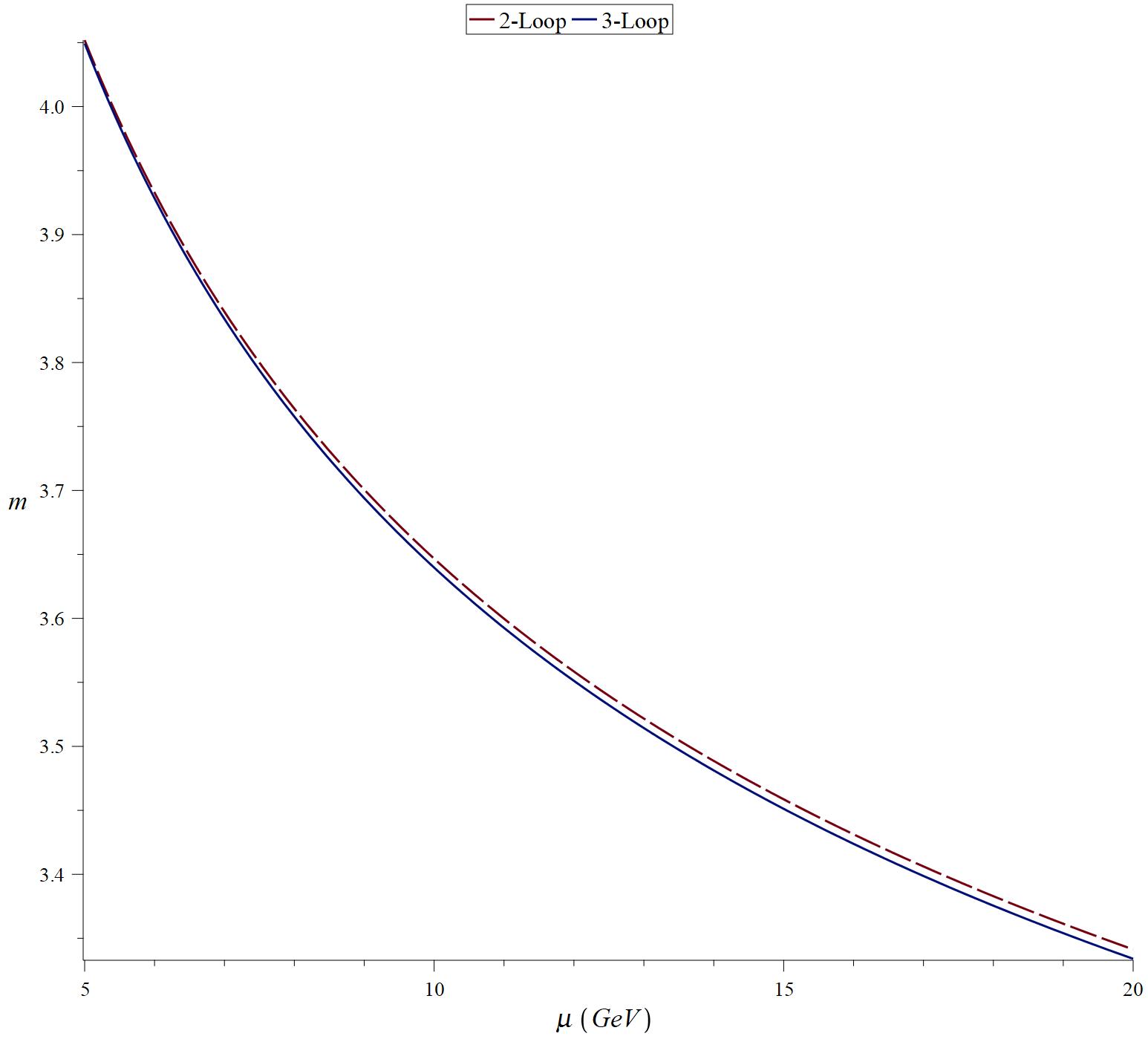}
\caption{The renormalization scale dependence of $b$-quark mass $m_b$ in the ${\overline{MS}}$ scheme at 2- and 3-loop orders}
\label{Fig. 2}
\end{center}
\end{figure}

\begin{figure}[hbt]
\begin{center}
\includegraphics[scale=0.36]{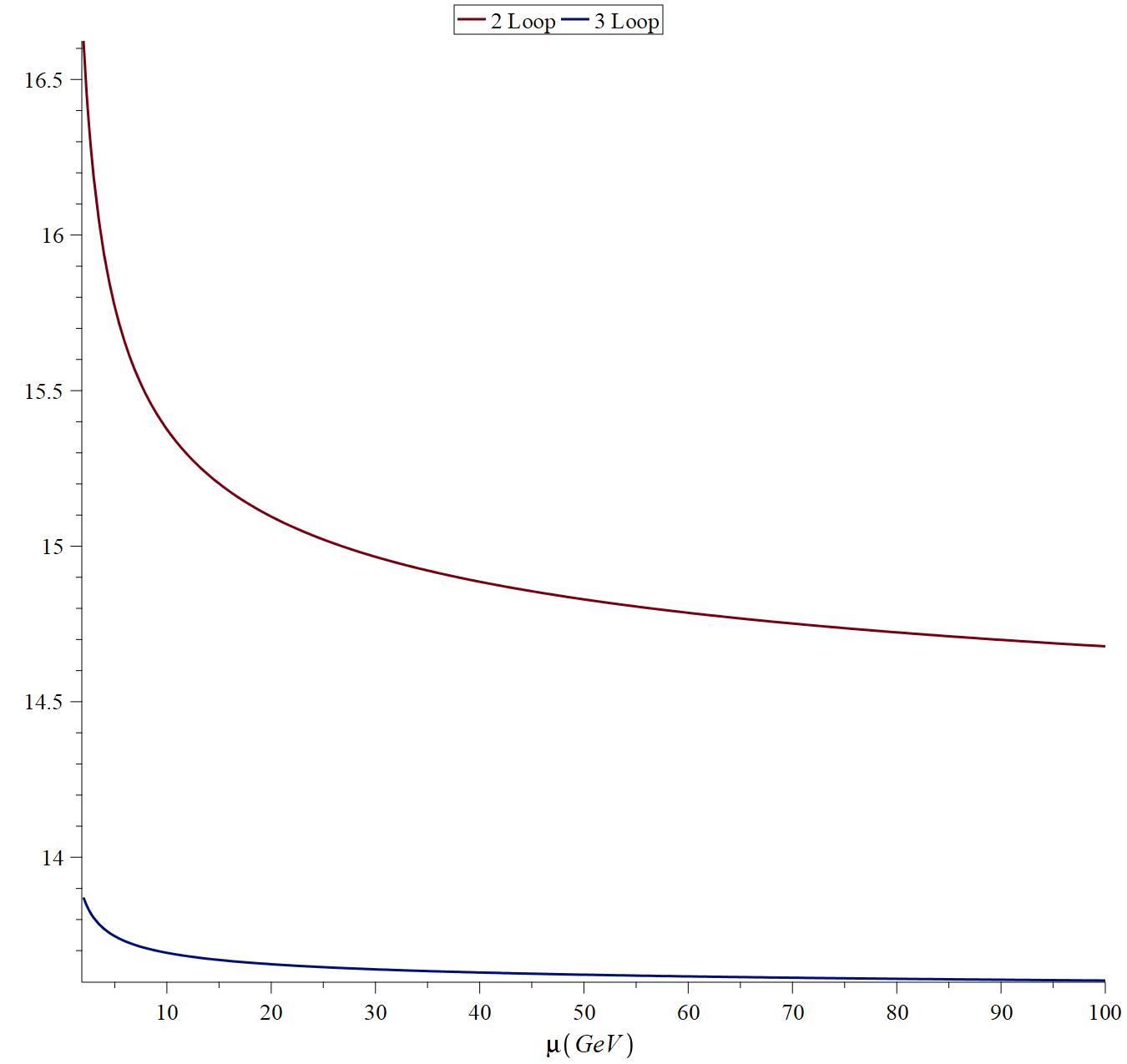}
\caption{The renormalization scale dependence of $I\!\!M$ at 2- and 3-loop orders}
\label{Fig. 3}
\end{center}
\end{figure}

\newpage

\begin{figure}[hbt]
\begin{center}
\includegraphics[scale=0.32]{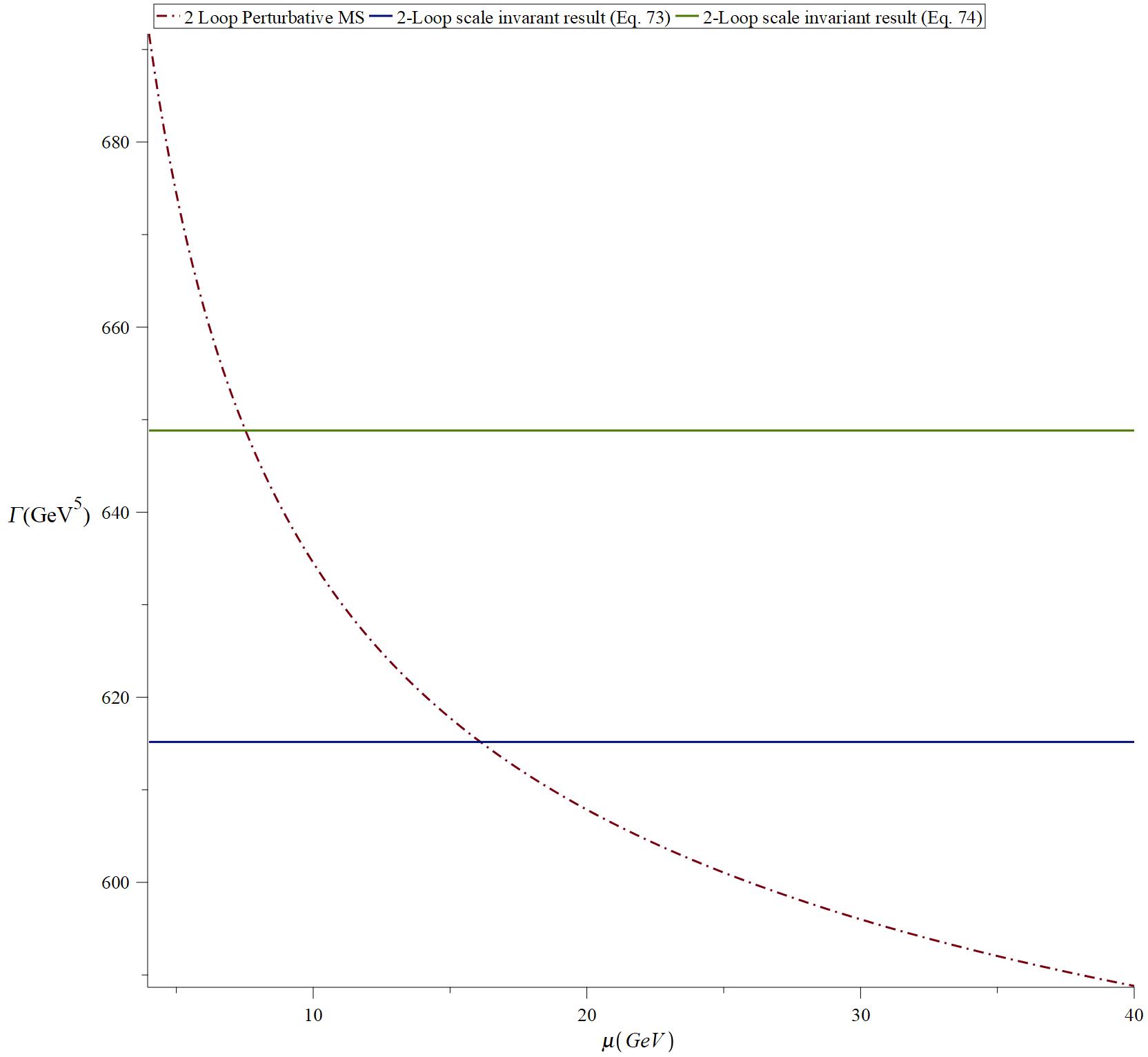}
\caption{The renormalization scale dependence of 2-loop perturbative $\Gamma$ as compared to the 2-Loop RG scale invariant results from Eqs. 73 and 74.}
\label{Fig. 4}
\end{center}
\end{figure}

\end{document}